\documentclass[english, jmp, reprint, amsmath, amssymb]{revtex4-2}
\usepackage[T1]{fontenc}
\usepackage[utf8]{inputenc}
\usepackage[table]{xcolor}
\usepackage{amssymb}
\usepackage{float}
\usepackage{graphicx}
\usepackage{babel}
\usepackage[normalem]{ulem}
\usepackage[detect-weight]{siunitx}
\graphicspath{ {./build/} }

\usepackage[unicode=true,pdfusetitle,
 bookmarks=true,bookmarksnumbered=true,bookmarksopen=true,bookmarksopenlevel=1,
 breaklinks=false,pdfborder={0 0 0},pdfborderstyle={}]{hyperref}
\usepackage[capitalize]{cleveref}

\begin{document}

\title{Josephson junctions based on ultraclean carbon nanotubes}

\author{S. Annabi$^{1}$}
\author{E. Arrighi$^{1}$, A. Peugeot$^{1}$, H. Riechert$^{1}$, J. Griesmar$^{1}$, K. Watanabe$^{2}$, T. Taniguchi$^{3}$}
\author{L. Bretheau$^{1*}$}
\author{J.-D. Pillet$^{1}$}
\selectlanguage{english}

\altaffiliation{These authors supervised equally this work.
\newline
landry.bretheau@polytechnique.edu 
\newline
jean-damien.pillet@polytechnique.edu}
\affiliation{$^{1}$Laboratoire de Physique de la Mati\`ere condens\'ee, CNRS, \'Ecole polytechnique, Institut Polytechnique de Paris, 91120 Palaiseau, France}
\affiliation{$^{2}$Research Center for Electronic and Optical Materials, National Institute for Materials Science, 1-1 Namiki, Tsukuba 305-0044, Japan}
\affiliation{$^{3}$Research Center for Materials Nanoarchitectonics, National Institute for Materials Science,  1-1 Namiki, Tsukuba 305-0044, Japan
}

\begin{abstract}
We present a technique for integrating ultraclean carbon nanotubes into superconducting circuits, aiming to realize Josephson junctions based on one-dimensional elementary quantum conductors. This technique primarily involves depositing the nanotube in the final step, thus preserving it from the inherent contaminations of nanofabrication and maintaining contact solely with superconducting electrodes and a crystalline hBN substrate. Through transport measurements performed in both the normal and superconducting states, we demonstrate that our method yields high-quality junctions with Josephson energies suitable for quantum device applications, such as carbon nanotube-based superconducting qubits.

\end{abstract}

\maketitle

\section{Introduction}

Carbon nanotubes (CNT) exhibit electronic behaviors akin to ideal one-dimensional nanowires, positioning them as promising materials for quantum conductors in modern electronic applications~\cite{laird_quantum_2015,baydin_carbon_2022}. Specifically, they can be utilized to confine a single electron, exploiting its charge, spin, or valley degrees of freedom for quantum information storage~\cite{laird_valleyspin_2013,pei_hyperfine_2017,penfold-fitch_microwave_2017,khivrich_atomic-like_2020}. When interfaced with superconducting electrodes, CNT serve as weak links, enabling the engineering of Josephson junctions with only a couple of spin-degenerate conduction channels~\cite{cleuziou_carbon_2006,jarillo-herrero_quantum_2006,pallecchi_carbon_2008} in the sense of Landauer~\cite{landauer_spatial_1957,akkermans_introduction_2007,nazarov_quantum_2009}. Such hybrid junctions offer a versatile platform for the realization of gate-tunable superconducting qubits, including gatemons and Andreev qubits, which have successfully been realized in atomic contact, semiconducting nanowires and two-dimensional electron gas~\cite{de_lange_realization_2015,larsen_semiconductor-nanowire-based_2015,casparis_superconducting_2018,kringhoj_magnetic-field-compatible_2021,wang_coherent_2019,janvier_coherent_2015,hays_coherent_2021,pita-vidal_direct_2023}. The intrinsic simplicity and limited internal degrees of freedom of CNT render them attractive candidates for mitigating decoherence mechanisms stemming from intrinsic relaxation or dephasing processes. Despite their potential, the coherent control of CNT-based circuits remains a challenging goal~\cite{mergenthaler_circuit_2021}. Pioneering efforts in this field have achieved only limited success, with fidelity and coherence times falling short of expectations. These limitations are predominantly attributed to disorder and noise in the systems.

In recent years, new techniques for nanoassembly of CNT have emerged~\cite{cao_electron_2005,deshpande_one-dimensional_2008,deshpande_mott_2009,wu_one-step_2010,Jung2013,waissman_realization_2013,jung_ultraclean_2013,baumgartner_Carbon_2014,huang_superior_2015,cheng_guiding_2019,cubaynes_nanoassembly_2020,lotfizadeh_quantum_2021,althuon_nano-assembled_2024} aiming to produce less disordered devices, particularly well-suited for qubit implementation. These techniques are mostly based on suspending the CNT. The resulting devices, referred to as ultraclean, preserve the CNT from substrate imperfections and residues left by previous nanofabrication steps. However, these new nanoassembly techniques are not easily compatible with the constraints imposed by modern architectures of quantum devices. One of the key challenges is to identify an approach suitable for the realization of superconducting qubits, which requires making junctions with large enough critical current~$I_\mathrm{c}$. The corresponding Josephson energies ${E_\mathrm{J}=\varphi_0 I_\mathrm{c}}$, with $\varphi_0$ the reduced flux quantum, must be on the order of several GHz when normalized by Planck's constant.

In this work, we present a novel approach for fabricating pristine Josephson junctions based on individual CNT. Akin to the revolutionary techniques recently developed for graphene and its derivatives~\cite{dean_boron_2010,wang_one-dimensional_2013}, we use crystalline hexagonal boron nitride (hBN) as a defect-free substrate. Our method involves the pickup of a suspended CNT using hBN and its deposition at the last step onto superconducting electrodes, keeping the CNT at a distance from the amorphous silicon oxide substrate (\cref{fig:fig_fab}a-c). 
A key strength of our approach is the quality of the electrical contacts, which is a crucial element for achieving critical currents compatible with the realization of superconducting qubits. We characterize our devices by performing transport measurements above and below the critical temperature of the superconducting electrodes. The low disorder level in our CNT-based circuits manifests in the regularity of the patterns observed in conductance and critical current as we sweep the gate voltage~$V_\mathrm{g}$ controlling the CNT's chemical potential. The remarkable quality of our devices demonstrate that our nanoassembly technique places CNT as promising quantum conductors that can serve as weak links for superconducting quantum devices. 

\section{CNT integration in superconducting circuits}

\begin{figure}
\includegraphics[width=1\columnwidth]{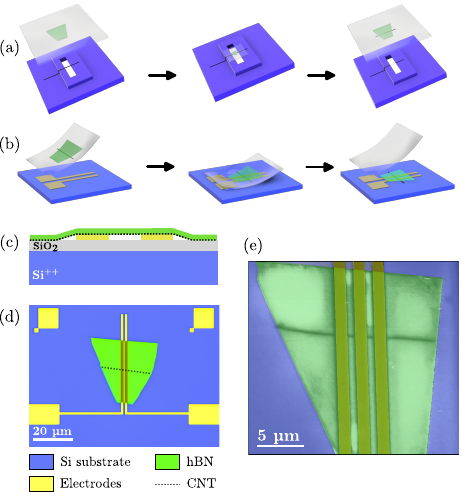}
\caption{(a) Pick-up steps for assembling the hBN-CNT stack. The PDMS is depicted in transparent grey, the hBN in green, the growth chip in purple and the CNT as a thin black line. (b) Integration of the CNT into the superconducting circuit. The substrate is depicted in blue, and the electrodes are in yellow. (c) Profile schematic of an assembled device with the same color code as in (d). (d) Optical image of a fully assembled device. The CNT is not visible with an optical microscope and is sketched here as a dashed black line for illustrative purposes. (e) False-colored atomic force microscope (AFM) image of the device labeled A, which was designed with three electrodes. The color code remains consistent with before. The CNT can be observed below the hBN by applying a voltage to the conductive tip of the AFM, appearing as the thick black line across the electrodes.}
\label{fig:fig_fab}
\end{figure}

We synthesize our CNT through a standard chemical vapor deposition (CVD) process on a silicon substrate~\cite{kong_chemical_1998,huang_growth_2003}. The substrate features a central slit above which CNT are suspended following growth (\cref{fig:fig_fab}a). This geometry, used in prior works~\cite{cheng_guiding_2019}, allows for optical detection and characterization of the nanotubes through Rayleigh spectroscopy~\cite{sfeir_probing_2004,huang_controlled_2005,sfeir_optical_2006}. It is consequently possible to selectively identify individual, single-walled CNT of known chirality (see \cref{appendix_rayleigh}). Such precise selection is crucial for electronic device design, as it allows to determine whether the CNT is metallic or semiconducting.

Once identified, the CNT is carefully picked up using a thin flake of hBN, previously exfoliated onto a transparent and flexible polydimethylsiloxane (PDMS) substrate (\cref{fig:fig_fab}a)~\cite{castellanos-gomez_deterministic_2014}. Using a micromanipulator, the hBN is brought into contact with the suspended part of the nanotube, where it adheres through Van der Waals interactions. These interactions are sufficiently strong that upon removal of the hBN sheet, the CNT breaks at its ends detaching from the silicon substrate. This results in an hBN-CNT stack, which is subsequently transferred onto a set of electrodes (\cref{fig:fig_fab}b), nanofabricated on a $300$ nm layer of SiO$_2$ covering a doped silicon substrate used as a back gate (see \cref{appendix_nanofab} for nanofabrication details).

After the transfer, the CNT comes into direct contact with the electrodes (\cref{fig:fig_fab}c, d and e). In most of our samples, these electrodes consist of a bilayer Nb(35nm)-Au(7nm), making them superconducting at low temperatures while avoiding oxidation when exposed to air. This is crucial for maintaining a good electrical connection with the nanotube. Importantly, the nanotube is not exposed to solvents or resist throughout the transfer process, avoiding the deposition of residues on or around its surface. Another significant advantage of this transfer technique is the suspension of the hBN-CNT stack between the electrodes, provided they are sufficiently close. This prevents the nanotube from directly contacting the amorphous silicon oxide of the substrate between the electrodes (see AFM profile in \cref{fig:devices_schematic}e), resulting in an ultraclean CNT-based Josephson junction. Notably, the integration process is carried out in ambient air making this technique easily accessible and highly practical for widespread implementation.

The quality of our devices was assessed by transport measurements, initially at room temperature to evaluate the transparency of the contacts between the CNT and the electrodes, followed by measurements at low temperatures in a dilution cryostat. These were performed at temperatures both above (\SI{3.6}{\kelvin}) and below (\SI{10}{\milli\kelvin}) the critical temperature $T_\mathrm{c}\approx\SI{3}{\kelvin}$ of the superconducting electrodes, allowing them to be in either their normal or superconducting states, respectively. In the rest of this manuscript, we focus on measurements performed on two of our best devices, exhibiting resistances of the order of a few tens of k$\Omega$ at room temperature (see \cref{appendix_yield} for detailed statistics). One device features a CNT identified as semiconducting (device~A), while the other is based on a pre-identifed metallic CNT (device~B). These devices were initially designed to function respectively as a SQUID and an Andreev molecule~\cite{pillet_nonlocal_2019,kornich_fine_2019,pillet_scattering_2020,matsuo_observation_2022,haxell_demonstration_2023,pillet_josephson_2023,keliri_driven_2023} for experiments beyond the scope of this manuscript, but for the purposes of the work presented here they were measured under conditions in which they behave as simple Josephson junctions (see \cref{appendix_architectures}).

\section{Transport measurements in the normal state ($T>T_\mathrm{c}$)}

\begin{figure}
\includegraphics[width=\columnwidth]{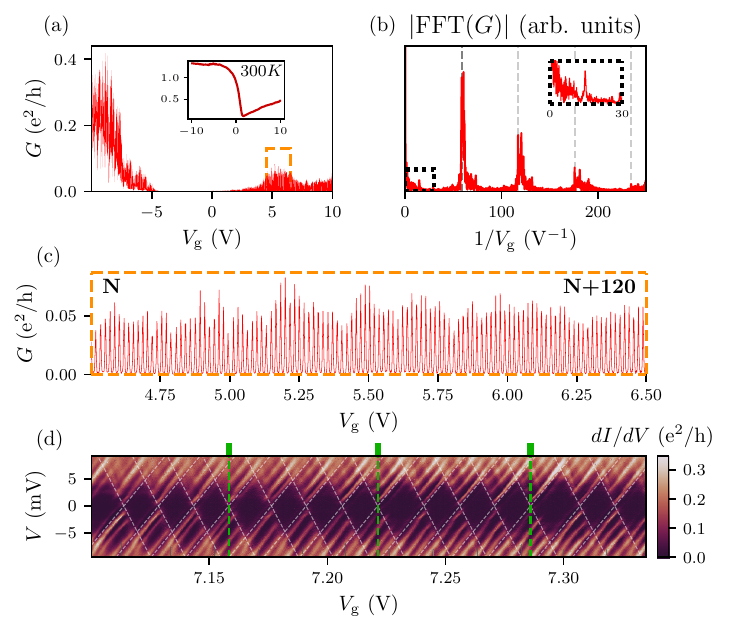}
\caption{
(a) Conductance of device~A measured at \SI{3.6}{\kelvin} plotted against gate voltage at zero bias. Inset: same measurement conducted at room temperature. (b) Fourier transform of the zero-bias conductance at positive gate voltage. The fundamental frequency is indicated by the first vertical dashed line, corresponding to a period of \SI{17.1}{\milli\volt}; other dashed lines highlight higher harmonics. Inset: close-up of the left portion of the graph revealing a low-frequency component exactly four times slower than the main period. (c) Zoomed-in view of the orange-highlighted region in (a) displaying more than 120 regularly spaced peaks. (d) Differential conductance plotted as a function of gate voltage and bias voltage, exhibiting Coulomb diamonds with a fourfold degeneracy highlighted by the green dashed lines. The heights of these diamonds yield the charging energy $U\approx\SI{5}{\milli\eV}$ and the confinement energy $\Delta E\approx\SI{1.5}{\milli\eV}$. Capacitance to the source, drain, and back gate, $C_s\approx \SI{13}{\atto\farad}$, $C_d\approx \SI{10}{\atto\farad}$, and $C_g\approx \SI{11}{\atto\farad}$, are extracted from the slopes of the white dashed lines defining the diamonds.}
\label{fig:fig_A}
\end{figure}

\begin{figure}
\includegraphics[width=\columnwidth]{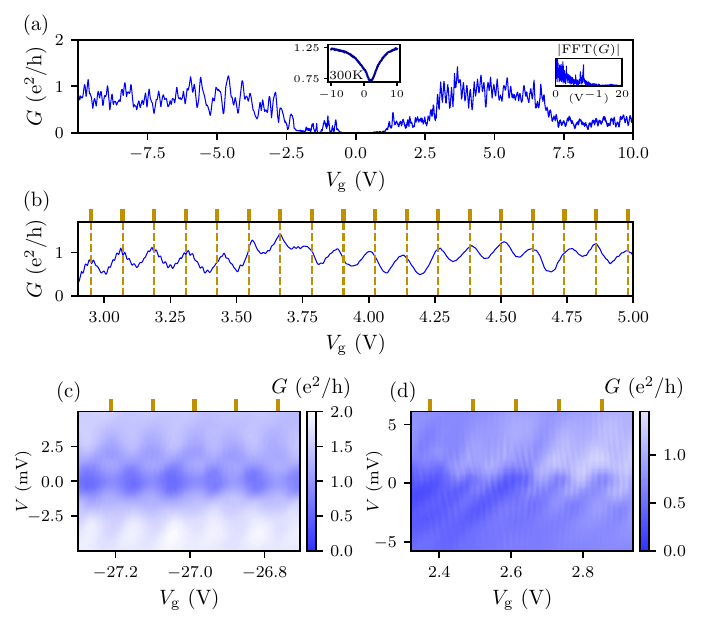}
\caption{
(a) Conductance of device~B measured at \SI{3.6}{\kelvin} plotted against gate voltage at zero bias. Middle inset: similar measurement conducted at room temperature. Top right inset: Fourier transform of the zero-bias conductance across the entire gate voltage range of (a). The dominant frequency component corresponds, this time, to a gate voltage period of \SI{120}{\milli\volt}. (b) Zoomed-in view of (a) at positive gate voltages displaying regular oscillations at the period identified in the Fourier transform, highlighted by dashed brown lines. (c)-(d) Conductance plotted as a function of gate voltage and bias voltage, at large negative gate voltages and small positive gate voltages. On top of the slow oscillations, we observe fast oscillations with a \SI{17}{\milli\volt} period.}
\label{fig:fig_B}
\end{figure}

As a first step, the CNT junctions are characterized with the electrodes turned in their normal state by measuring at a temperature $T$ above their critical temperature $T_\mathrm{c}\approx\SI{3}{\kelvin}$. Electronic transport measurements performed on the semiconducting and metallic devices, A and B respectively, are presented in \cref{fig:fig_A} and \cref{fig:fig_B}. These initial measurements offer a clear and direct assessment of the low disorder level in our devices.

At room temperature ($T\gg T_\mathrm{c}$), device~A exhibits a conductance minimum around $V_\mathrm{g} = \SI{0}{\volt}$ (inset of \cref{fig:fig_A}a), corresponding to the Fermi energy lying in the middle of the semiconducting gap.  
Away from this operating point, the conductance increases with $|V_\mathrm{g}|$, with an observed asymmetry between positive and negative voltages associated with hole doping by the gold-covered electrodes~\cite{Heinze2002}. 
At low temperature ($T=\SI{3.6}{\kelvin}>T_\mathrm{c}$), the device is in the Coulomb blockade regime, a typical behavior of standard CNT junctions~\cite{Tans1997}. This regime is evidenced by the emergence of hundreds of conductance peaks with regular spacing (\cref{fig:fig_A}c), indicative of high cleanliness~\cite{contamin_zero_2022}, as confirmed by the Fourier transform of the signal (\cref{fig:fig_A}b). When applying an additional bias voltage across the junction, we observe a Coulomb diamond pattern (\cref{fig:fig_A}d) with a regularity that corroborates the pristine nature of the device. 

A more detailed examination reveals a fourfold periodic pattern reminiscent of the spin and valley degeneracy~\cite{minot_determination_2004}, as revealed by the inset of \cref{fig:fig_A}b. Specifically, every fourth diamond appears slightly larger, with respective heights of $6.5$~meV and $5$~meV. While the smaller value is simply the charging energy $U$ of the device, the additional $1.5$~meV corresponds to the confinement energy $h v_F/(2L)$, where $h$ is Planck's constant, $v_F\sim \SI{8.1e5}{\meter\per\second}$ the Fermi velocity of the nanotube, and $L$ the length of the CNT between the electrodes, \emph{i.e.} \SI{1}{\micro\meter} in our design. The observation of the four-fold degeneracy provides a widely acknowledged benchmark of the exceptional cleanliness of the device~\cite{liang_shell_2002,cao_electron_2005,makarovski_su2_2007,kuemmeth_coupling_2008,grove-rasmussen_superconductivity-enhanced_2009,jespersen_gate-dependent_2011,cleuziou_interplay_2013,delagrange_0-ensuremathpi_2016,althuon_nano-assembled_2024}. Indeed, even minimal disorder tends to couple the clockwise and counterclockwise orbital motion of the electrons around the nanotube, breaking and eliminating the fourfold degeneracy.

Device B, fabricated with a metallic nanotube, also exhibits a minimum conductance near $V_\mathrm{g}=0$ V at room temperature (inset of \cref{fig:fig_B}a). As the temperature decreases, this minimum evolves to nearly approaching zero in a certain gate voltage window. This phenomenon can be explained by the Dirac point or the opening of an energy gap, although the precise mechanism remains to be identified~\cite{ouyang_energy_2001,deshpande_mott_2009,senger_universal_2018, Hu2024}. In line with expectations for a metallic CNT, the conductance of this device exceeds that of the previous one, reaching values close to the conductance quantum $e^2/h$. Upon sweeping the gate voltage, Coulomb blockade peaks are not observed but rather oscillations arising from electronic interferences (\cref{fig:fig_B}b). These oscillations turn into a checkerboard pattern (\cref{fig:fig_B}c and d), which is a generally acknowledged signature of an electronic Fabry-Pérot interferometer~\cite{liang_fabry_2001,herrmann_shot_2007,yang_fabry-perot_2020,lotfizadeh_quantum_2021}. This observation provides compelling evidence of a low level of disorder in our device.

\section{Critical current of CNT-based Josephson junctions}

\begin{figure}
\includegraphics[width=\columnwidth]{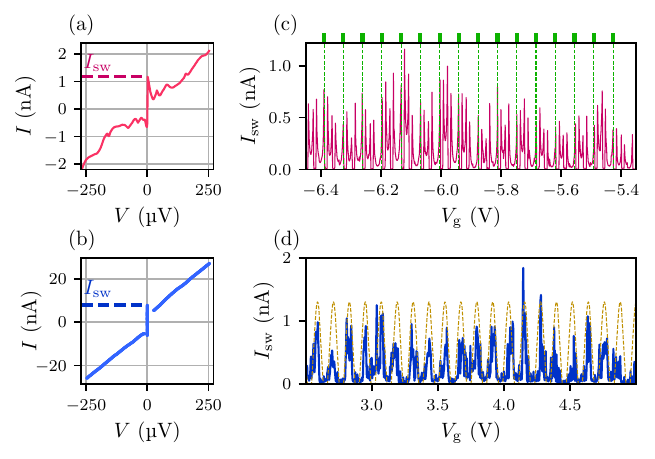}
\caption{(a)-(b) Current-voltage characteristics of devices A (a) and B (b) measured at \SI{10}{\milli\kelvin}, with respective gate voltages of \SI{-6.1212}{\volt} and \SI{-12.794}{\volt}. (c)-(d) Plot of $I_\mathrm{sw}$ as a function of gate voltage for devices A (c) and B (d). The green dashed lines in (c) indicate the periodic repetition of a four-peak pattern every \SI{64}{\milli\volt} stemming from the fourfold degeneracy of the CNT. In (d), the brown dashed line represents a cosine function with a period of \SI{120}{\milli\volt}, corresponding to the Fabry-Pérot periodicity extracted in the normal state.}
\label{fig:fig_Isw}
\end{figure}

As we lower the temperature of our devices to \SI{10}{\milli\kelvin} well below $T_\mathrm{c}$, the electrodes transition into their superconducting states, leading to the observation of various regimes in the device, such as the mere opening of a superconducting gap or the emergence of Andreev states~\cite{pillet_andreev_2010,pillet_tunneling_2011,pillet_tunneling_2013} (see \cref{appendix_supra}). Here, our focus is solely on the gate voltage range where the Josephson effect is measurable. It manifests as a supercurrent branch in the current-voltage characteristic, as illustrated in \cref{fig:fig_Isw}a and b. The maximum of this supercurrent branch, known as the switching current $I_\mathrm{sw}$, exhibits a strong dependence on the gate voltage. Owing to finite noise on the junction, $I_\mathrm{sw}$ is in practice lower than the intrinsic critical current $I_\mathrm{c}$ of the device~\cite{vion_thermal_1996,jorgensen_critical_2007,eichler_tuning_2009}, the latter determining the Josephson energy $E_\mathrm{J}$ (see appendix \ref{appendix_Isw}). Switching currents were measured up to \SI{4}{\nano\ampere} in device~A and \SI{8}{\nano\ampere} in device~B, corresponding to $E_\mathrm{J}/h$ of at least \SI{2}{\giga\hertz} and \SI{4}{\giga\hertz} respectively, which is suitable for the implementation of gatemon qubits. Both metallic and semiconducting nanotubes are thus compatible with the realization of superconducting qubits.

Figures \ref{fig:fig_Isw}c and d show the evolution of $I_{\text{sw}}$ with $V_\mathrm{g}$. Device A exhibits a series of peaks reminiscent of the Coulomb oscillations discussed earlier, along with a recurring pattern corresponding to the fourfold degeneracy of the nanotube. While a detailed examination of this pattern and its underlying physical factors exceeds the scope of this manuscript and will be addressed in future work, it once again underscores the quality of our devices, this time through their superconducting properties. Device B exhibits higher switching currents consistent with the metallic nature of the nanotube. It also presents a somewhat erratic evolution with gate voltage, the exact cause of which remains unclear. We hypothesize that the proximity of a second junction in our designs may perturb the behavior of the first one, potentially through the hybridization of the Andreev states~\cite{pillet_nonlocal_2019,kornich_fine_2019}. Alternatively, this behavior could stem from an enhanced sensitivity of the switching current to disorder, revealing nuances not readily apparent in measurements conducted in the normal state. Nonetheless, these irregular oscillations manifest in periodic clusters (highlighted by the dashed brown curve in \cref{fig:fig_Isw}d), which coincide with the period of Fabry-Pérot oscillations observed in the normal state. This alignment suggests that the supercurrent is conveyed by a ballistic conductor, thereby affirming the ultraclean nature of the nanotube.

\section{Concluding remarks}

We have demonstrated that our approach of integrating CNT onto superconducting electrodes enables the fabrication of Josephson junctions based on unique ultraclean nanotubes. Our observations from transport measurements hold promise for future microwave architectures, such as superconducting qubits or quantum sensors \cite{cleuziou_carbon_2006, khivrich_atomic-like_2020}. Utilizing Josephson junctions with minimal internal electronic degrees of freedom is expected to mitigate sources of decoherence such as dephasing due to quasiparticle poisoning \cite{Bretheau_supercurrent_spectro_2013}. Furthermore, our devices comprising CNT shielded by a layer of hBN, exhibit notable durability and resilience against electrostatic discharges and high electrical currents. Crucially, this nanoassembly technique is compatible with Circuit Quantum Electrodynamics architectures, provided that intrinsic silicon, rather than doped silicon, is used for the substrate to minimize dissipation, along with a local gate~\cite{larsen_semiconductor-nanowire-based_2015, de_lange_realization_2015, wang_coherent_2019} designed atop the hBN flake. Additionally, we have observed gradual improvements in contact quality over time in a majority of our devices, with contacts becoming increasingly transparent, possibly due to the gradual removal of microbubbles between the hBN layer and the electrodes. Although the reproducibility level of nanotube junctions has not yet reached that of Josephson tunnel junctions, nanotubes nonetheless offer a promising alternative for future quantum devices.

More generally, this nanoassembly technique has a great potential in term of versatility. For instance, the nature of the electrodes onto which the nanotube is transferred could be substituted with other superconductors than niobium, or even extend to normal and ferromagnetic electrodes, facilitating the study of exotic hybrid systems~\cite{feuillet-palma_conserved_2010,herrmann_carbon_2010,desjardins_synthetic_2019,bordoloi_spin_2022,wang_singlet_2022}. Additionally, the potential substitution of hBN with other two-dimensional materials, particularly transition metal dichalcogenides (TMDs) with a similar lattice parameter as graphene, could lead to inducing strong spin-orbit interactions in the nanotube~\cite{masseroni_spin-orbit_2024}. This could pave the way towards the realization of complex one-dimensional architectures such as Kitaev chains~\cite{dvir_realization_2023}, Andreev polymers~\cite{pillet_nonlocal_2019,johannsen_fermionic_2024} or 1D moir\'e systems~\cite{zhou_pressure_2024}. Another exciting direction would be to harness nanotubes as charge sensors for exotic 2D quantum materials~\cite{cheng_guiding_2019}.

\begin{acknowledgements}
We acknowledge valuable discussions with S. Park and F. Cadiz on hBN exfoliation and transfer. Thanks to S. Delacroix, L. Virey, J. Sulpizio, M. Desjardins, C. Jadaud, and A. Cheng for their guidance on Rayleigh scattering and CVD growth of carbon nanotubes. We also thank P. Kim and L. Anderson for providing growth chips used for preliminary tests. Special thanks go to A. Vecchiola for his assistance with AFM imaging, as well as R. Mohammedi and D. Roux for their technical support. Gratitude is extended to R. Ribeiro-Palau, F. Parmentier, P.F. Orfila, S. Delprat, M.F. Goffman, D. Vion, and the SPEC of CEA-Saclay for their help on nanofabrication processes. We appreciate the insights of Ç. Girit, J.-L. Smirr, and L. Peyruchat on cryogenic and low-noise measurements. Additionally, we acknowledge the expertise shared by I. Aupiais, R. Grasset, and Y. Laplace on optical setups. Special thanks are given to M. Delbecq, T. Kontos, S. Guéron, R. Deblock, the EPS group, and the Quantronics group for engaging scientific discussions. JDP acknowledges support from the Agence Nationale de la Recherche (grant ANR-20-CE47-0003), and LB from the European Research Council (grant agreement No. 947707). Financial support from Ecole Polytechnique and the Region Ile-de-France within the framework of DIM SIRTEQ is also acknowledged by JDP and LB. This work has been supported by the French ANR-22-PETQ-0003 grant under the France 2030 plan. K.W. and T.T. acknowledge support from the JSPS KAKENHI (Grant Numbers 21H05233 and 23H02052) and World Premier International Research Center Initiative (WPI), MEXT, Japan.
\end{acknowledgements}

\begin{appendix}
\renewcommand{\thesubsection}{APPENDIX \Alph{subsection}:}

\section{Nanofabrication of the superconducting electrodes}
\label{appendix_nanofab}

The electrodes onto which the carbon nanotubes are transferred are nanofabricated by e-beam deposition of metal through a resist mask. Typically, we deposit a metallic bilayer of \SI{35}{\nano\meter} of niobium covered by \SI{7}{\nano\meter} of gold. This choice is guided by numerical calculations of the expected critical temperature $T_\mathrm{c}$ and AFM imaging confirming that the gold film is continuous for this thickness. The mask is created by electron beam lithography on a bilayer of MAA-PMMA resist to increase the undercut and ensure that the deposited metal is not in direct contact with the resist. The resulting surface of the electrodes is cleaner and suitable for the transfer of CNT.

\section{Yield of CNT integration}
\label{appendix_yield}

\begin{figure}
\includegraphics[width=\columnwidth]{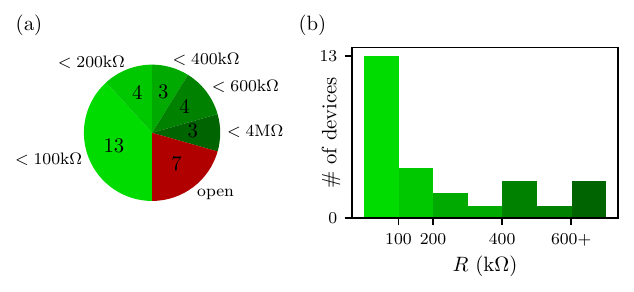}
\caption{(a) Distribution of resistances among the 34 fabricated devices. (b) Distribution of resistances among the 27 devices having a measurable resistance at room temperature.}
\label{fig:Statistics}
\end{figure}

We evaluate the performance of our nanoassembly technique by calculating the yield of the CNT integration step on metallic and superconducting electrodes. This step involves picking up the nanotube with the hBN flake and depositing the hBN-CNT stack onto the electrodes. Diagnosis is performed by measuring the minimum of the device resistance $R$, immediately after transfer, as we sweep the gate voltage.

It is not possible to use the Ambegaokar and Baratoff formula, which applies only to tunnel junctions, to deduce the critical current from the normal-state resistance measured at room temperature $R$. However, in the limit of a large gap ${\Delta \gg \Gamma}$, where $\Gamma$ represents the coupling of the nanotube to the electrodes, one can relate the normal-state resistance to the critical current via the expression ${I_\mathrm{c} \sim \pi\Delta_\mathrm{eff} / (2eR)}$, where $\Delta_\mathrm{eff}$ is an effective superconducting gap~\cite{Meng_self_consistent_2009}. This approximation is reasonable in our case, as $\Delta \approx \SI{450}{\micro\eV}$ and $\Gamma$ is typically of the order of~\SI{100}{\micro\eV} for nanotubes~\cite{Maurand_First_2012}. We can then distinguish between two regimes: the Coulomb blockade regime, where~${U \gg \Gamma}$, as in device~A, and the Fabry-Perot regime, where ${U \ll \Gamma}$, as in device~B. In these cases, the effective gap is ${\Delta_\mathrm{eff} = \Gamma^2 / U}$ and $\Gamma$, respectively. However, these parameters are not known a priori during the room-temperature testing. For reference, the critical current of a device with a normal state resistance of \SI{100}{\kilo\ohm} is of the order of \SI{100}{\pico\ampere} if the device operates in the Coulomb blockade regime with a typical charging energy of \SI{1}{\milli\eV}, and of the order of \SI{1}{\nano\ampere} if it operates in the Fabry-Pérot regime.

Statistics of successful assembly is shown in \cref{fig:Statistics}. Out of the 34 assembled samples, 27 exhibit a finite resistance $R$ at room temperature. Failed attempts were mostly due to breaking the CNT during pick-up, or the entrapment of immovable air bubbles between the CNT and the electrode after the deposition step. Consequently, the success rate is $80\%$. More importantly, half of the successfully assembled devices exhibited resistances below \SI{100}{\kilo\ohm}, which is the upper limit for achieving a sufficiently large Josephson energy. Among these, 9 were identified as semiconducting by Rayleigh scattering, and 4 as metallic.

Six devices were measured at low temperature (\SI{10}{\milli\kelvin}), most of them demonstrating similar regularity to the devices presented in this manuscript. Importantly, for devices made with superconducting electrodes, the switching current reaches values exceeding \SI{300}{\pico\ampere} in $75\%$ of cases. For some devices, the contacts became increasingly transparent after storing them for a few months under \SI{1}{\milli\bar} vacuum, which is attributed to the gradual removal of microbubbles between the hBN layer and the electrodes. In practice, their resistance decreased and the measured switching current increased by an order of magnitude. In the case of device~A, the room-temperature resistance decreased from \SI{70}{\kilo\ohm} to \SI{20}{\kilo\ohm} at ${V_\mathrm{g}=\SI{-10}{\volt}}$, and the maximum measured switching current increased from \SI{300}{\pico\ampere} to \SI{4}{\nano\ampere}.

\section{Rayleigh spectroscopy}
\label{appendix_rayleigh}

\begin{figure}
\includegraphics[width=\columnwidth]{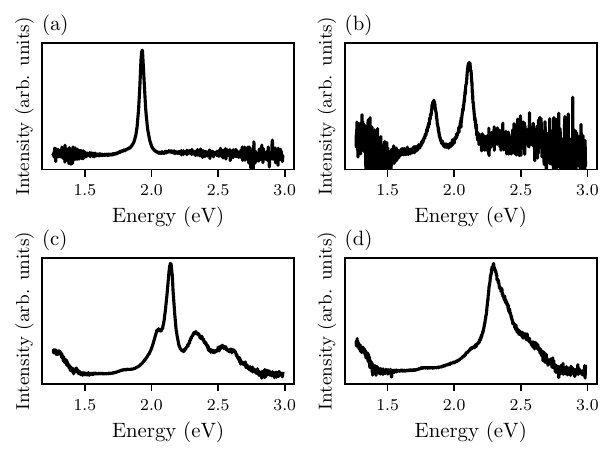}
\caption{Rayleigh spectra of four different CNT. (a) Spectrum of a metallic armchair nanotube of chirality (10,10). (b) Spectrum of a semi-conducting nanotube of chirality (19,11). (c) Spectrum of the CNT used for device~B, identified as a metallic nanotube of chirality (22,16). (d) Spectrum of the CNT used for device~A, identified as a semi-conducting nanotube of chirality (13,9).}
\label{fig:Rayleigh}
\end{figure}

For CNT characterization, we perform Rayleigh spectroscopy. It is a photoluminescence technique that involves focusing a broadband laser on the CNT and collecting the re-emitted light. The signal is mostly composed of elastically scattered light, which is then analyzed using a spectrometer. The resulting spectrum displays peaks, which energies correspond to transitions between van Hove singularities in the density of states of the CNT~\cite{sfeir_probing_2004}, providing a fingerprint for identifying the chirality~\cite{kataura_optical_1999,liu_atlas_2012}.

\cref{fig:Rayleigh} shows 4 examples of typical spectra obtained by Rayleigh spectroscopy. When only a few peaks are present in our observation window, between \SI{1.25} and \SI{3}{\eV}, it is possible to perform chirality identification with good reliability. In some cases, an additional background is observed, as in c and d, likely due to spurious scattered light, either from impurities on the nanotube, the silicon substrate, or another nearby nanotube.

\section{Measurement setup}

\begin{figure}
\includegraphics[width=\columnwidth]{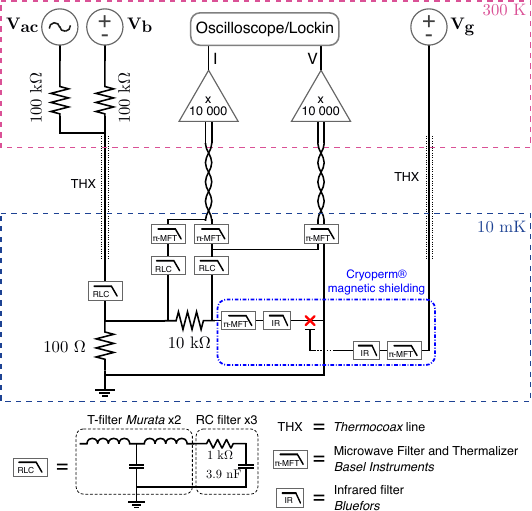}
\caption{Schematic of the experimental setup for transport measurements. The junction, represented by a red cross, is capacitively coupled to the gate. Gate and bias voltages $V_\mathrm{g}$ and $V_\mathrm{b}$ are controlled with a DC voltage source, and an additional alternative bias voltage $V_{ac}$ is applied with a lockin amplifier for differential conductance measurements. All the lines are filtered with different types of filters placed at \SI{10}{\milli\kelvin}.}
\label{fig:measurement_setup}
\end{figure}

\cref{fig:measurement_setup} presents the experimental apparatus used to perform the transport measurements described in this work. A voltage $V_\mathrm{b}$ is applied at room temperature using a low-noise voltage source \textit{Yokogawa GS200}. It is divided by 1000 at low temperatures by a voltage divider to improve the voltage precision on the junction as well as divide the technical noise from the instrument. A \SI{10}{\kilo\ohm} resistor is in series with the device in order to measure the current through it. Current and voltage signals are measured via resistive twisted pairs and amplified by low noise differential voltage amplifiers from Basel Instruments. In order to measure a current-voltage characteristic, the bias voltage $V_\mathrm{b}$ is ramped on the desired range with a typical ramp time of 3 seconds, while current and voltage are measured with an oscilloscope. Differential conductance signals can be obtained by taking the numerical derivative of such a measurement, or directly measured with a lockin-amplifier. The gate voltage is applied using a low-noise voltage source \textit{Basel Instruments SP927}.

All the lines are filtered to reduce the electronic noise on the device and increase the measured switching current \cite{vion_thermal_1996,jorgensen_critical_2007,eichler_tuning_2009}. Four types of filters are used to cover the whole noise spectrum. The junction is directly connected on on side to the \SI{10}{\milli\kelvin} ground. The bias and the gate electrodes are filtered with infrared filters (\textit{Bluefors}) shielding the junction from high-energy photons, and with microwave filters from \textit{Basel Instruments} attenuating signals above \SI{10}{\mega\hertz}. A combination of T-filters and RC filters are used to reduce noise above \SI{10}{\kilo\hertz}. Additionally, \textit{Thermocoax} lines are used to apply the bias voltage as well as gate voltage, effectively filtering microwave noise \cite{Zorin1998}. All these filters are cryogenic and placed at the lowest stage of our dilution cryostat to optimize their performance.

\section{Switching current versus Josephson energy}
\label{appendix_Isw}

\begin{figure}
\includegraphics[width=\columnwidth]{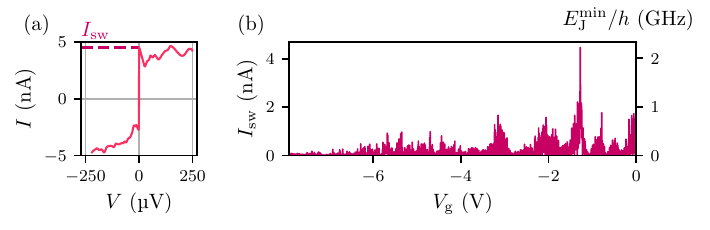}
\caption{(a) Current-voltage characteristic measured in device~A with $I_\mathrm{sw}=\SI{4.5}{\nano\ampere}$, which is the largest value we observed for this device. (b) Same measurements of $I_\mathrm{sw}$ as in \cref{fig:fig_Isw}c but on a larger gate voltage scale. On the right axis, $I_\mathrm{sw}$ is converted in a lower estimate $E_\mathrm{J}^\mathrm{min}$ of the Josephson energy normalized by Planck's constant~$h$.}
\label{I_sw_vs_E_J}
\end{figure}

The difference between the switching current ($I_\mathrm{sw}$) and the critical current ($I_\mathrm{c}$) can be quite substantial. This discrepancy arises from the noise in the environment surrounding the junction, and depends on the value of $I_\mathrm{c}$. Consequently, there is no straightforward relation linking $I_\mathrm{sw}$ to $I_\mathrm{c}$. For a device like ours, which has a low critical current (i.e., $E_\mathrm{J}$ on the order of a few $k_\mathrm{B} T_\mathrm{e}$, where $k_\mathrm{B}$ is the Boltzmann constant and $T_\mathrm{e}$ is the electronic temperature) and a large parallel capacitance due to the back gate (estimated at around \SI{30}{\pico\farad}), the junction operates in the thermally assisted regime or phase diffusion regime~\cite{lemasne_asymmetric_2010}. In such devices, the supercurrent branch may experimentally not be visible, or it may show a supercurrent peak with a maximum that is up to ten times lower than the intrinsic value of $I_\mathrm{c}$~\cite{eichler_tuning_2009,jorgensen_critical_2007}, meaning ${I_\mathrm{sw} \sim \qtyrange[range-phrase=-~, range-units=single]{100}{200}{\pico\ampere}}$ for ${I_\mathrm{c} \sim \qtyrange[range-phrase=-~, range-units=single]{1}{2}{\nano\ampere}}$.

Numerical simulations, based on a classical description of the circuit~\cite{Goffman_Supercurrent_2000,cron_these_2001} with parameters similar to those in our experiment, indicate a significant reduction of $I_\mathrm{sw}$ relative to $I_\mathrm{c}$. For instance, a critical current of 10, 2, and \SI{1}{\nano\ampere} would approximately yield switching currents of \SI{6}{\nano\ampere}, \SI{400}{\pico\ampere}, and \SI{100}{\pico\ampere}, respectively. Thus, our measurements of $I_\mathrm{sw}$ provide a conservative estimate of $E_\mathrm{J}$ (see Fig. \ref{I_sw_vs_E_J}), which further reinforces the conclusion that our junctions are suitable for implementing a superconducting qubit, such as a nanotube gatemon.

\section{Architectures of device A and B}
\label{appendix_architectures}

\begin{figure}
\includegraphics[width=\columnwidth]{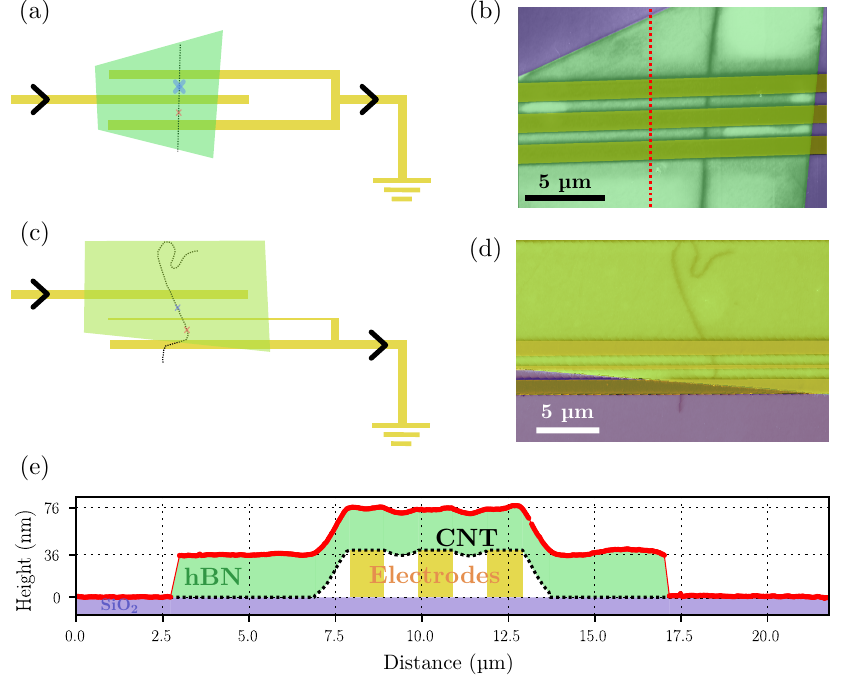}
\caption{(a) Sketch of device~A layout in a SQUID configuration. The electrodes are represented in yellow, the hBN in green and the CNT in dashed black line forming two junctions in parallel indicated by the blue and red crosses. Current flows as indicated by the black arrows. (b) Corresponding false-colored AFM image; the silicon dioxide layer is shown in purple. (c) Sketch of device~B layout in an Andreev molecule configuration (same color code). (d) Corresponding false-colored AFM image. (e) Topographic profile of device  A along a cut indicated by the red line in (b). The profile (in red) remains flat between two electrodes, suggesting that the hBN-CNT stack is suspended between the electrodes.}
\label{fig:devices_schematic}
\end{figure}

Device A consists of 2 junctions in parallel (\cref{fig:devices_schematic}a and b) forming a SQUID. However, we observed that one of the two junctions (marked with a red cross on the schematic) did not contribute to the supercurrent, likely because of less transparent electrical contacts. In this regime, the device behaves like a single Josephson junction (marked with a blue cross on the schematic), without the periodic flux dependence of the critical current that should be observed in a SQUID. The data presented in this manuscript were obtained under these conditions. Only for large values of gate voltage does the second Josephson junction carry a significant portion of the supercurrent. This regime is not covered in this manuscript.

Device B, shown in~\cref{fig:devices_schematic}c and d, was initially designed to conduct an experiment on Andreev molecules~\cite{pillet_nonlocal_2019}. It consists of a single Josephson junction (blue cross on the schematic) placed nearby a second one (red cross on the schematic) that is short-circuited by a loop in order to control the superconducting phase difference across it by applying a magnetic flux through it. The main objective of this device is to demonstrate non-local Josephson effect, a physical phenomenon that we do not address in this manuscript, and no magnetic flux was applied in the measurements shown in this work. Nonetheless, this architecture allows for the measurement of the current-voltage characteristic of only one of the two junctions. While the behavior of this junction might subtly be influenced by the presence of the second one, it primarily behaves as a single Josephson junction.

\section{Additional transport measurements in the normal state ($T>T_\mathrm{c}$)}
\label{appendix_normal}

\begin{figure}
\includegraphics[width=\columnwidth]{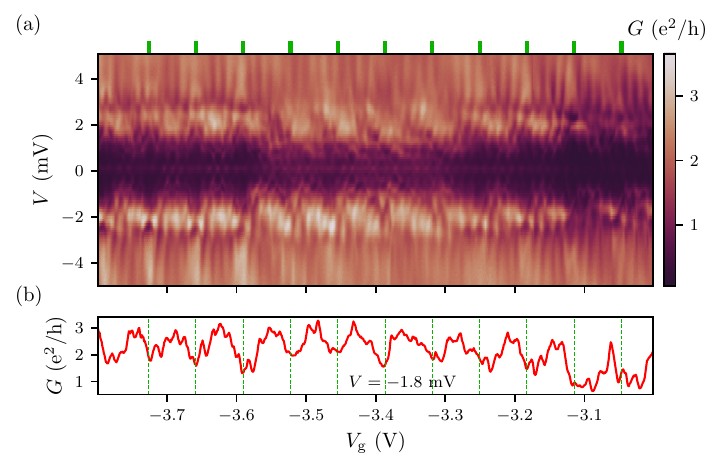}
\caption{Differential conductance of device~A measured at \SI{3.6}{\kelvin} as a function of gate voltage and bias voltage at negative gate voltages.}
\label{fig:fig_NTS_negative_gate}
\end{figure}

\cref{fig:fig_NTS_negative_gate} shows conductance measurements of Device A conducted in the normal state (${T=\SI{3.6}{\kelvin}}$), with the nanotube electrostatically doped with holes (${V_\mathrm{g}<0}$), in contrast to \cref{fig:fig_A}c and d in the main text where the nanotube was doped with electrons (${V_\mathrm{g}>0}$). On the hole side, the conductance is significantly higher and Coulomb blockade is barely visible. Nevertheless, the fourfold periodicity, that we use as a benchmark for cleanliness, remains discernible even for negative gate voltages.

\section{Additional transport measurements in the superconducting state ($T<T_\mathrm{c}$)}
\label{appendix_supra}

\begin{figure}
\includegraphics[width=\columnwidth]{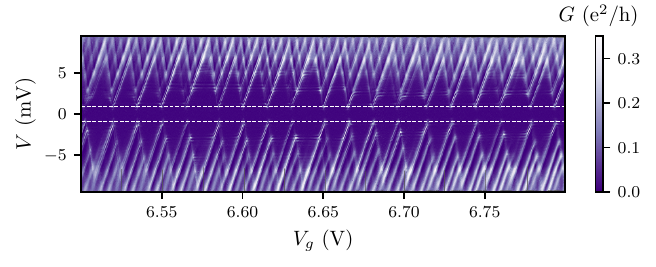}
\caption{Differential conductance of device~A measured at \SI{10}{\milli\kelvin} as a function of gate voltage and bias voltage. The electrodes' superconducting gap is highlighted by the horizontal white dashed lines.}
\label{fig:fig_A_coulomb_supra}
\end{figure}

Device A exhibits Coulomb blockade at positive gate voltages when the electrodes are in their metallic state (\cref{fig:fig_A}). \cref{fig:fig_A_coulomb_supra} shows similar measurements but performed at the base temperature of the cryostat, such that the electrodes are in their superconducting state. In this regime, the device does not carry any sizeable supercurrent and the sole effect of superconductivity is simply to split the Coulomb diamonds around zero bias voltage and open a superconducting gap $\Delta$. We can then estimate the latter ${\Delta\approx\SI{450}{\micro\eV}}$, which is consistent with the critical temperature of ${T_\mathrm{c}\approx\SI{3}{\kelvin}}$ of our Nb-Au bilayer~\cite{tinkham_introduction_1996}.

\begin{figure}
\includegraphics[width=\columnwidth]{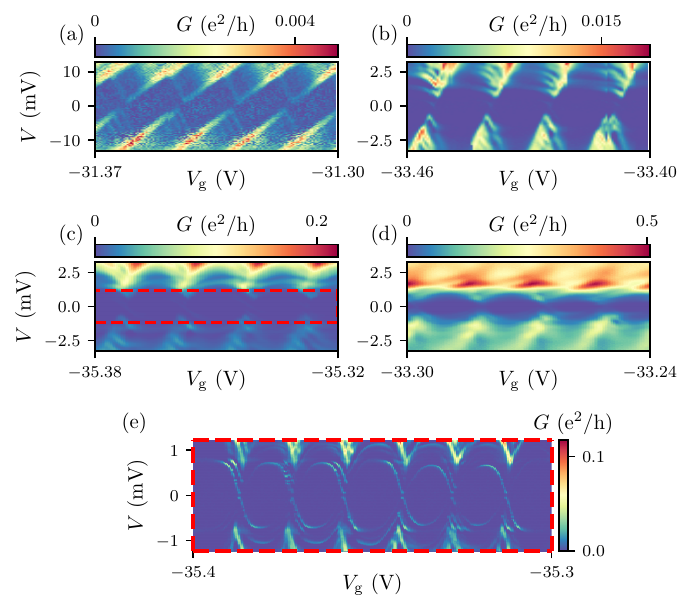}
\caption{Numerical differential conductance of device C measured at \SI{10}{\milli\kelvin} as a function of gate voltage and bias voltage, in different gate voltage regions. (e) Close-up of (c) inside the superconducting gap.}
\label{fig:JJNT}
\end{figure}

\begin{figure}
\includegraphics[width=\columnwidth]{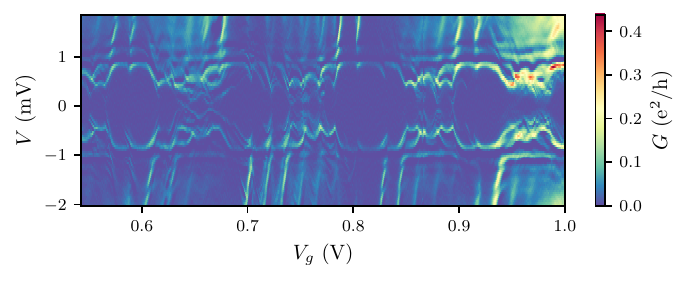}
\caption{Conductance measurements performed in device~B at ${T\sim \SI{10}{\milli\kelvin}}$. Subgap resonances suggest the presence of Andreev Bound States.}
\label{fig:NTM_spiders}
\end{figure}

\cref{fig:JJNT} presents additional transport measurements performed at low temperature on another device, labeled C. This device is composed of a semiconducting nanotube connected to two superconducting electrodes with a similar structure to devices A and B, forming a single Josephson junction. Around $V_\mathrm{g}\sim \SI{-31.3}{\volt}$, we observe Coulomb diamonds split by a superconducting gap similarly to device~A. By sweeping the gate voltage to lower values, we drive device C in regimes of larger conductance. The sharply defined Coulomb diamonds tend to be smoothed out and we observe additional subgap features, highlighted in \cref{fig:JJNT}e and resembling Andreev bound states~\cite{pillet_andreev_2010}. This behavior was also observed in device~B for small gate voltages (\cref{fig:NTM_spiders}).

\end{appendix}

\newpage
\bibliography{Nanofab_CNT}

\begin{thebibliography}{10}

\bibitem{laird_quantum_2015}
E.~A. Laird, F.~Kuemmeth, G.~A. Steele, K.~Grove-Rasmussen, J.~Nyg{\r a}rd, K.~Flensberg, and L.~P. Kouwenhoven, ``Quantum transport in carbon nanotubes,'' {\em Reviews of Modern Physics}, vol.~87, pp.~703--764, July 2015.
\newblock Publisher: American Physical Society.

\bibitem{baydin_carbon_2022}
A.~Baydin, F.~Tay, J.~Fan, M.~Manjappa, W.~Gao, and J.~Kono, ``Carbon {Nanotube} {Devices} for {Quantum} {Technology},'' {\em Materials}, vol.~15, p.~1535, Feb. 2022.

\bibitem{laird_valleyspin_2013}
E.~A. Laird, F.~Pei, and L.~P. Kouwenhoven, ``A valley{\textendash}spin qubit in a carbon nanotube,'' {\em Nature Nanotechnology}, vol.~8, pp.~565--568, Aug. 2013.
\newblock Publisher: Nature Publishing Group.

\bibitem{pei_hyperfine_2017}
T.~Pei, A.~P{\'a}lyi, M.~Mergenthaler, N.~Ares, A.~Mavalankar, J.~H. Warner, G.~A.~D. Briggs, and E.~A. Laird, ``Hyperfine and {Spin}-{Orbit} {Coupling} {Effects} on {Decay} of {Spin}-{Valley} {States} in a {Carbon} {Nanotube},'' {\em Physical Review Letters}, vol.~118, p.~177701, Apr. 2017.
\newblock Publisher: American Physical Society.

\bibitem{penfold-fitch_microwave_2017}
Z.~V. Penfold-Fitch, F.~Sfigakis, and M.~R. Buitelaar, ``Microwave {Spectroscopy} of a {Carbon} {Nanotube} {Charge} {Qubit},'' {\em Physical Review Applied}, vol.~7, p.~054017, May 2017.
\newblock Publisher: American Physical Society.

\bibitem{khivrich_atomic-like_2020}
I.~Khivrich and S.~Ilani, ``Atomic-like charge qubit in a carbon nanotube enabling electric and magnetic field nano-sensing,'' {\em Nature Communications}, vol.~11, p.~2299, May 2020.
\newblock Publisher: Nature Publishing Group.

\bibitem{cleuziou_carbon_2006}
J.-P. Cleuziou, W.~Wernsdorfer, V.~Bouchiat, T.~Ondar{\c c}uhu, and M.~Monthioux, ``Carbon nanotube superconducting quantum interference device,'' {\em Nature Nanotechnology}, vol.~1, pp.~53--59, Oct. 2006.
\newblock Publisher: Nature Publishing Group.

\bibitem{jarillo-herrero_quantum_2006}
P.~Jarillo-Herrero, J.~A. van Dam, and L.~P. Kouwenhoven, ``Quantum supercurrent transistors in carbon nanotubes,'' {\em Nature}, vol.~439, pp.~953--956, Feb. 2006.
\newblock Publisher: Nature Publishing Group.

\bibitem{pallecchi_carbon_2008}
E.~Pallecchi, M.~Gaa{\ss}, D.~A. Ryndyk, and C.~Strunk, ``Carbon nanotube {Josephson} junctions with {Nb} contacts,'' {\em Applied Physics Letters}, vol.~93, p.~072501, Aug. 2008.

\bibitem{landauer_spatial_1957}
R.~Landauer, ``Spatial {Variation} of {Currents} and {Fields} {Due} to {Localized} {Scatterers} in {Metallic} {Conduction},'' {\em IBM Journal of Research and Development}, vol.~1, pp.~223--231, July 1957.
\newblock Conference Name: IBM Journal of Research and Development.

\bibitem{akkermans_introduction_2007}
``Introduction: mesoscopic physics,'' in {\em Mesoscopic {Physics} of {Electrons} and {Photons}} (E.~Akkermans and G.~Montambaux, eds.), pp.~1--30, Cambridge: Cambridge University Press, 2007.

\bibitem{nazarov_quantum_2009}
Y.~V. Nazarov and Y.~M. Blanter, {\em Quantum {Transport}: {Introduction} to {Nanoscience}}.
\newblock Cambridge: Cambridge University Press, 2009.

\bibitem{de_lange_realization_2015}
G.~de~Lange, B.~van Heck, A.~Bruno, D.~J. van Woerkom, A.~Geresdi, S.~R. Plissard, E.~P. A.~M. Bakkers, A.~R. Akhmerov, and L.~DiCarlo, ``Realization of {Microwave} {Quantum} {Circuits} {Using} {Hybrid} {Superconducting}-{Semiconducting} {Nanowire} {Josephson} {Elements},'' {\em Physical Review Letters}, vol.~115, p.~127002, Sept. 2015.
\newblock Publisher: American Physical Society.

\bibitem{larsen_semiconductor-nanowire-based_2015}
T.~W. Larsen, K.~D. Petersson, F.~Kuemmeth, T.~S. Jespersen, P.~Krogstrup, J.~Nyg{\r a}rd, and C.~M. Marcus, ``Semiconductor-{Nanowire}-{Based} {Superconducting} {Qubit},'' {\em Physical Review Letters}, vol.~115, p.~127001, Sept. 2015.
\newblock Publisher: American Physical Society.

\bibitem{casparis_superconducting_2018}
L.~Casparis, M.~R. Connolly, M.~Kjaergaard, N.~J. Pearson, A.~Kringh{\o}j, T.~W. Larsen, F.~Kuemmeth, T.~Wang, C.~Thomas, S.~Gronin, G.~C. Gardner, M.~J. Manfra, C.~M. Marcus, and K.~D. Petersson, ``Superconducting gatemon qubit based on a proximitized two-dimensional electron gas,'' {\em Nature Nanotechnology}, vol.~13, pp.~915--919, Oct. 2018.
\newblock Number: 10 Publisher: Nature Publishing Group.

\bibitem{kringhoj_magnetic-field-compatible_2021}
A.~Kringh{\o}j, T.~W. Larsen, O.~Erlandsson, W.~Uilhoorn, J.~Kroll, M.~Hesselberg, R.~McNeil, P.~Krogstrup, L.~Casparis, C.~Marcus, and K.~Petersson, ``Magnetic-{Field}-{Compatible} {Superconducting} {Transmon} {Qubit},'' {\em Physical Review Applied}, vol.~15, p.~054001, May 2021.
\newblock Publisher: American Physical Society.

\bibitem{wang_coherent_2019}
J.~I.-J. Wang, D.~Rodan-Legrain, L.~Bretheau, D.~L. Campbell, B.~Kannan, D.~Kim, M.~Kjaergaard, P.~Krantz, G.~O. Samach, F.~Yan, J.~L. Yoder, K.~Watanabe, T.~Taniguchi, T.~P. Orlando, S.~Gustavsson, P.~Jarillo-Herrero, and W.~D. Oliver, ``Coherent control of a hybrid superconducting circuit made with graphene-based van der {Waals} heterostructures,'' {\em Nature Nanotechnology}, vol.~14, pp.~120--125, Feb. 2019.
\newblock Number: 2 Publisher: Nature Publishing Group.

\bibitem{janvier_coherent_2015}
C.~Janvier, L.~Tosi, L.~Bretheau, {\c C}.~{\"O}. Girit, M.~Stern, P.~Bertet, P.~Joyez, D.~Vion, D.~Esteve, M.~F. Goffman, H.~Pothier, and C.~Urbina, ``Coherent manipulation of {Andreev} states in superconducting atomic contacts,'' {\em Science}, vol.~349, pp.~1199--1202, Sept. 2015.
\newblock Publisher: American Association for the Advancement of Science.

\bibitem{hays_coherent_2021}
M.~Hays, V.~Fatemi, D.~Bouman, J.~Cerrillo, S.~Diamond, K.~Serniak, T.~Connolly, P.~Krogstrup, J.~Nyg{\r a}rd, A.~Levy~Yeyati, A.~Geresdi, and M.~H. Devoret, ``Coherent manipulation of an {Andreev} spin qubit,'' {\em Science}, vol.~373, pp.~430--433, July 2021.
\newblock Publisher: American Association for the Advancement of Science.

\bibitem{pita-vidal_direct_2023}
M.~Pita-Vidal, A.~Bargerbos, R.~{\v Z}itko, L.~J. Splitthoff, L.~Gr{\"u}nhaupt, J.~J. Wesdorp, Y.~Liu, L.~P. Kouwenhoven, R.~Aguado, B.~van Heck, A.~Kou, and C.~K. Andersen, ``Direct manipulation of a superconducting spin qubit strongly coupled to a transmon qubit,'' {\em Nature Physics}, vol.~19, pp.~1110--1115, Aug. 2023.
\newblock Publisher: Nature Publishing Group.

\bibitem{mergenthaler_circuit_2021}
M.~Mergenthaler, A.~Nersisyan, A.~Patterson, M.~Esposito, A.~Baumgartner, C.~Sch{\"o}nenberger, G.~A.~D. Briggs, E.~A. Laird, and P.~J. Leek, ``Circuit {Quantum} {Electrodynamics} with {Carbon}-{Nanotube}-{Based} {Superconducting} {Quantum} {Circuits},'' {\em Physical Review Applied}, vol.~15, p.~064050, June 2021.
\newblock Publisher: American Physical Society.

\bibitem{cao_electron_2005}
J.~Cao, Q.~Wang, and H.~Dai, ``Electron transport in very clean, as-grown suspended carbon nanotubes,'' {\em Nature Materials}, vol.~4, pp.~745--749, Oct. 2005.
\newblock Number: 10 Publisher: Nature Publishing Group.

\bibitem{deshpande_one-dimensional_2008}
V.~V. Deshpande and M.~Bockrath, ``The one-dimensional {Wigner} crystal in carbon nanotubes,'' {\em Nature Physics}, vol.~4, pp.~314--318, Apr. 2008.
\newblock Publisher: Nature Publishing Group.

\bibitem{deshpande_mott_2009}
V.~V. Deshpande, B.~Chandra, R.~Caldwell, D.~S. Novikov, J.~Hone, and M.~Bockrath, ``Mott {Insulating} {State} in {Ultraclean} {Carbon} {Nanotubes},'' {\em Science}, vol.~323, pp.~106--110, Jan. 2009.
\newblock Publisher: American Association for the Advancement of Science.

\bibitem{wu_one-step_2010}
C.~C. Wu, C.~H. Liu, and Z.~Zhong, ``One-{Step} {Direct} {Transfer} of {Pristine} {Single}-{Walled} {Carbon} {Nanotubes} for {Functional} {Nanoelectronics},'' {\em Nano Letters}, vol.~10, pp.~1032--1036, Mar. 2010.
\newblock Publisher: American Chemical Society.

\bibitem{Jung2013}
M.~Jung, J.~Schindele, S.~Nau, M.~Weiss, A.~Baumgartner, and C.~Sch{\"{o}}nenberger, ``{Ultraclean Single, Double, and Triple Carbon Nanotube Quantum Dots with Recessed Re Bottom Gates},'' {\em Nano Letters}, vol.~13, pp.~4522--4526, sep 2013.

\bibitem{waissman_realization_2013}
J.~Waissman, M.~Honig, S.~Pecker, A.~Benyamini, A.~Hamo, and S.~Ilani, ``Realization of pristine and locally tunable one-dimensional electron systems in carbon nanotubes,'' {\em Nature Nanotechnology}, vol.~8, pp.~569--574, Aug. 2013.
\newblock Number: 8 Publisher: Nature Publishing Group.

\bibitem{jung_ultraclean_2013}
M.~Jung, J.~Schindele, S.~Nau, M.~Weiss, A.~Baumgartner, and C.~Sch{\"o}nenberger, ``Ultraclean {Single}, {Double}, and {Triple} {Carbon} {Nanotube} {Quantum} {Dots} with {Recessed} {Re} {Bottom} {Gates},'' {\em Nano Letters}, vol.~13, pp.~4522--4526, Sept. 2013.
\newblock Publisher: American Chemical Society.

\bibitem{baumgartner_Carbon_2014}
A.~Baumgartner, G.~Abulizi, K.~Watanabe, T.~Taniguchi, J.~Gramich, and C.~Schönenberger, ``{Carbon nanotube quantum dots on hexagonal boron nitride},'' {\em Applied Physics Letters}, vol.~105, p.~023111, 07 2014.

\bibitem{huang_superior_2015}
J.-W. Huang, C.~Pan, S.~Tran, B.~Cheng, K.~Watanabe, T.~Taniguchi, C.~N. Lau, and M.~Bockrath, ``Superior {Current} {Carrying} {Capacity} of {Boron} {Nitride} {Encapsulated} {Carbon} {Nanotubes} with {Zero}-{Dimensional} {Contacts},'' {\em Nano Letters}, vol.~15, pp.~6836--6840, Oct. 2015.
\newblock Publisher: American Chemical Society.

\bibitem{cheng_guiding_2019}
A.~Cheng, T.~Taniguchi, K.~Watanabe, P.~Kim, and J.-D. Pillet, ``Guiding {Dirac} {Fermions} in {Graphene} with a {Carbon} {Nanotube},'' {\em Physical Review Letters}, vol.~123, p.~216804, Nov. 2019.
\newblock Publisher: American Physical Society.

\bibitem{cubaynes_nanoassembly_2020}
T.~Cubaynes, L.~C. Contamin, M.~C. Dartiailh, M.~M. Desjardins, A.~Cottet, M.~R. Delbecq, and T.~Kontos, ``Nanoassembly technique of carbon nanotubes for hybrid circuit-{QED},'' {\em Applied Physics Letters}, vol.~117, p.~114001, Sept. 2020.

\bibitem{lotfizadeh_quantum_2021}
N.~Lotfizadeh, M.~J. Senger, D.~R. McCulley, E.~D. Minot, and V.~V. Deshpande, ``Quantum {Interferences} in {Ultraclean} {Carbon} {Nanotubes},'' {\em Physical Review Letters}, vol.~126, p.~216802, May 2021.
\newblock Publisher: American Physical Society.

\bibitem{althuon_nano-assembled_2024}
T.~Althuon, T.~Cubaynes, A.~Auer, C.~S{\"u}rgers, and W.~Wernsdorfer, ``Nano-assembled open quantum dot nanotube devices,'' {\em Communications Materials}, vol.~5, pp.~1--7, Jan. 2024.
\newblock Publisher: Nature Publishing Group.

\bibitem{dean_boron_2010}
C.~R. Dean, A.~F. Young, I.~Meric, C.~Lee, L.~Wang, S.~Sorgenfrei, K.~Watanabe, T.~Taniguchi, P.~Kim, K.~L. Shepard, and J.~Hone, ``Boron nitride substrates for high-quality graphene electronics,'' {\em Nature Nanotechnology}, vol.~5, pp.~722--726, Oct. 2010.
\newblock Publisher: Nature Publishing Group.

\bibitem{wang_one-dimensional_2013}
L.~Wang, I.~Meric, P.~Y. Huang, Q.~Gao, Y.~Gao, H.~Tran, T.~Taniguchi, K.~Watanabe, L.~M. Campos, D.~A. Muller, J.~Guo, P.~Kim, J.~Hone, K.~L. Shepard, and C.~R. Dean, ``One-{Dimensional} {Electrical} {Contact} to a {Two}-{Dimensional} {Material},'' {\em Science}, vol.~342, pp.~614--617, Nov. 2013.
\newblock Publisher: American Association for the Advancement of Science.

\bibitem{kong_chemical_1998}
J.~Kong, A.~M. Cassell, and H.~Dai, ``Chemical vapor deposition of methane for single-walled carbon nanotubes,'' {\em Chemical Physics Letters}, vol.~292, pp.~567--574, Aug. 1998.

\bibitem{huang_growth_2003}
S.~Huang, X.~Cai, and J.~Liu, ``Growth of {Millimeter}-{Long} and {Horizontally} {Aligned} {Single}-{Walled} {Carbon} {Nanotubes} on {Flat} {Substrates},'' {\em Journal of the American Chemical Society}, vol.~125, pp.~5636--5637, May 2003.
\newblock Publisher: American Chemical Society.

\bibitem{sfeir_probing_2004}
M.~Y. Sfeir, F.~Wang, L.~Huang, C.-C. Chuang, J.~Hone, S.~P. O'Brien, T.~F. Heinz, and L.~E. Brus, ``Probing {Electronic} {Transitions} in {Individual} {Carbon} {Nanotubes} by {Rayleigh} {Scattering},'' {\em Science}, vol.~306, pp.~1540--1543, Nov. 2004.

\bibitem{huang_controlled_2005}
X.~M.~H. Huang, R.~Caldwell, L.~Huang, S.~C. Jun, M.~Huang, M.~Y. Sfeir, S.~P. O'Brien, and J.~Hone, ``Controlled {Placement} of {Individual} {Carbon} {Nanotubes},'' {\em Nano Letters}, vol.~5, pp.~1515--1518, July 2005.

\bibitem{sfeir_optical_2006}
M.~Y. Sfeir, T.~Beetz, F.~Wang, L.~Huang, X.~M.~H. Huang, M.~Huang, J.~Hone, S.~O'Brien, J.~A. Misewich, T.~F. Heinz, L.~Wu, Y.~Zhu, and L.~E. Brus, ``Optical {Spectroscopy} of {Individual} {Single}-{Walled} {Carbon} {Nanotubes} of {Defined} {Chiral} {Structure},'' {\em Science}, vol.~312, pp.~554--556, Apr. 2006.

\bibitem{castellanos-gomez_deterministic_2014}
A.~Castellanos-Gomez, M.~Buscema, R.~Molenaar, V.~Singh, L.~Janssen, H.~S. J. v.~d. Zant, and G.~A. Steele, ``Deterministic transfer of two-dimensional materials by all-dry viscoelastic stamping,'' {\em 2D Materials}, vol.~1, p.~011002, Apr. 2014.
\newblock Publisher: IOP Publishing.

\bibitem{pillet_nonlocal_2019}
J.-D. Pillet, V.~Benzoni, J.~Griesmar, J.-L. Smirr, and {\c C}.~{\"O}. Girit, ``Nonlocal {Josephson} {Effect} in {Andreev} {Molecules},'' {\em Nano Letters}, vol.~19, pp.~7138--7143, Oct. 2019.
\newblock Publisher: American Chemical Society.

\bibitem{kornich_fine_2019}
V.~Kornich, H.~S. Barakov, and Y.~V. Nazarov, ``Fine energy splitting of overlapping {Andreev} bound states in multiterminal superconducting nanostructures,'' {\em Physical Review Research}, vol.~1, p.~033004, Oct. 2019.
\newblock Publisher: American Physical Society.

\bibitem{pillet_scattering_2020}
J.-D. Pillet, V.~Benzoni, J.~Griesmar, J.-L. Smirr, and {\c C}.~Girit, ``Scattering description of {Andreev} molecules,'' {\em SciPost Physics Core}, vol.~2, p.~009, June 2020.

\bibitem{matsuo_observation_2022}
S.~Matsuo, J.~S. Lee, C.-Y. Chang, Y.~Sato, K.~Ueda, C.~J. Palmstr{\o}m, and S.~Tarucha, ``Observation of nonlocal {Josephson} effect on double {InAs} nanowires,'' {\em Communications Physics}, vol.~5, pp.~1--6, Sept. 2022.
\newblock Number: 1 Publisher: Nature Publishing Group.

\bibitem{haxell_demonstration_2023}
D.~Z. Haxell, M.~Coraiola, M.~Hinderling, S.~C. ten Kate, D.~Sabonis, A.~E. Svetogorov, W.~Belzig, E.~Cheah, F.~Krizek, R.~Schott, W.~Wegscheider, and F.~Nichele, ``Demonstration of the {Nonlocal} {Josephson} {Effect} in {Andreev} {Molecules},'' {\em Nano Letters}, vol.~23, pp.~7532--7538, Aug. 2023.
\newblock Publisher: American Chemical Society.

\bibitem{pillet_josephson_2023}
J.-D. Pillet, S.~Annabi, A.~Peugeot, H.~Riechert, E.~Arrighi, J.~Griesmar, and L.~Bretheau, ``Josephson diode effect in andreev molecules,'' {\em Phys. Rev. Res.}, vol.~5, p.~033199, Sep 2023.

\bibitem{keliri_driven_2023}
A.~Keliri and B.~Dou{\c c}ot, ``Driven {Andreev} molecule,'' {\em Physical Review B}, vol.~107, p.~094505, Mar. 2023.
\newblock Publisher: American Physical Society.

\bibitem{Heinze2002}
S.~Heinze, J.~Tersoff, R.~Martel, V.~Derycke, J.~Appenzeller, and P.~Avouris, ``{Carbon Nanotubes as Schottky Barrier Transistors},'' {\em Physical Review Letters}, vol.~89, p.~106801, aug 2002.

\bibitem{Tans1997}
S.~J. Tans, M.~H. Devoret, H.~Dai, A.~Thess, R.~E. Smalley, L.~J. Geerligs, and C.~Dekker, ``{Individual single-wall carbon nanotubes as quantum wires},'' {\em Nature}, vol.~386, pp.~474--477, apr 1997.

\bibitem{contamin_zero_2022}
L.~C. Contamin, L.~Jarjat, W.~Legrand, A.~Cottet, T.~Kontos, and M.~R. Delbecq, ``Zero energy states clustering in an elemental nanowire coupled to a superconductor,'' {\em Nature Communications}, vol.~13, p.~6188, Oct. 2022.
\newblock Number: 1 Publisher: Nature Publishing Group.

\bibitem{minot_determination_2004}
E.~D. Minot, Y.~Yaish, V.~Sazonova, and P.~L. McEuen, ``Determination of electron orbital magnetic moments in carbon nanotubes,'' {\em Nature}, vol.~428, pp.~536--539, Apr. 2004.
\newblock Publisher: Nature Publishing Group.

\bibitem{liang_shell_2002}
W.~Liang, M.~Bockrath, and H.~Park, ``Shell {Filling} and {Exchange} {Coupling} in {Metallic} {Single}-{Walled} {Carbon} {Nanotubes},'' {\em Physical Review Letters}, vol.~88, p.~126801, Mar. 2002.
\newblock Publisher: American Physical Society.

\bibitem{makarovski_su2_2007}
A.~Makarovski, A.~Zhukov, J.~Liu, and G.~Finkelstein, ``{SU}(2) and {SU}(4) {Kondo} effects in carbon nanotube quantum dots,'' {\em Physical Review B}, vol.~75, p.~241407, June 2007.
\newblock Publisher: American Physical Society.

\bibitem{kuemmeth_coupling_2008}
F.~Kuemmeth, S.~Ilani, D.~C. Ralph, and P.~L. McEuen, ``Coupling of spin and orbital motion of electrons in carbon nanotubes,'' {\em Nature}, vol.~452, pp.~448--452, Mar. 2008.
\newblock Publisher: Nature Publishing Group.

\bibitem{grove-rasmussen_superconductivity-enhanced_2009}
K.~Grove-Rasmussen, H.~I. J{\o}rgensen, B.~M. Andersen, J.~Paaske, T.~S. Jespersen, J.~Nyg{\r a}rd, K.~Flensberg, and P.~E. Lindelof, ``Superconductivity-enhanced bias spectroscopy in carbon nanotube quantum dots,'' {\em Physical Review B}, vol.~79, p.~134518, Apr. 2009.
\newblock Publisher: American Physical Society.

\bibitem{jespersen_gate-dependent_2011}
T.~S. Jespersen, K.~Grove-Rasmussen, J.~Paaske, K.~Muraki, T.~Fujisawa, J.~Nyg{\r a}rd, and K.~Flensberg, ``Gate-dependent spin{\textendash}orbit coupling in multielectron carbon nanotubes,'' {\em Nature Physics}, vol.~7, pp.~348--353, Apr. 2011.
\newblock Publisher: Nature Publishing Group.

\bibitem{cleuziou_interplay_2013}
J.~P. Cleuziou, N.~V. N{\textquoteright}Guyen, S.~Florens, and W.~Wernsdorfer, ``Interplay of the {Kondo} {Effect} and {Strong} {Spin}-{Orbit} {Coupling} in {Multihole} {Ultraclean} {Carbon} {Nanotubes},'' {\em Physical Review Letters}, vol.~111, p.~136803, Sept. 2013.
\newblock Publisher: American Physical Society.

\bibitem{delagrange_0-ensuremathpi_2016}
R.~Delagrange, R.~Weil, A.~Kasumov, M.~Ferrier, H.~Bouchiat, and R.~Deblock, ``0-pi quantum transition in a carbon nanotube {Josephson} junction: {Universal} phase dependence and orbital degeneracy,'' {\em Physical Review B}, vol.~93, p.~195437, May 2016.
\newblock Publisher: American Physical Society.

\bibitem{ouyang_energy_2001}
M.~Ouyang, J.-L. Huang, C.~L. Cheung, and C.~M. Lieber, ``Energy {Gaps} in "{Metallic}" {Single}-{Walled} {Carbon} {Nanotubes},'' {\em Science}, vol.~292, pp.~702--705, Apr. 2001.
\newblock Publisher: American Association for the Advancement of Science.

\bibitem{senger_universal_2018}
M.~J. Senger, D.~R. McCulley, N.~Lotfizadeh, V.~V. Deshpande, and E.~D. Minot, ``Universal interaction-driven gap in metallic carbon nanotubes,'' {\em Physical Review B}, vol.~97, p.~035445, Jan. 2018.
\newblock Publisher: American Physical Society.

\bibitem{Hu2024}
C.~Hu, J.~Chen, X.~Zhou, Y.~Xie, X.~Huang, Z.~Wu, S.~Ma, Z.~Zhang, K.~Xu, N.~Wan, Y.~Zhang, Q.~Liang, and Z.~Shi, ``{Collapse of carbon nanotubes due to local high-pressure from van der Waals encapsulation},'' {\em Nature Communications}, vol.~15, p.~3486, apr 2024.

\bibitem{liang_fabry_2001}
W.~Liang, M.~Bockrath, D.~Bozovic, J.~H. Hafner, M.~Tinkham, and H.~Park, ``Fabry - {Perot} interference in a nanotube electron waveguide,'' {\em Nature}, vol.~411, pp.~665--669, June 2001.
\newblock Number: 6838 Publisher: Nature Publishing Group.

\bibitem{herrmann_shot_2007}
L.~G. Herrmann, T.~Delattre, P.~Morfin, J.-M. Berroir, B.~Pla{\c c}ais, D.~C. Glattli, and T.~Kontos, ``Shot {Noise} in {Fabry}-{Pérot} {Interferometers} {Based} on {Carbon} {Nanotubes},'' {\em Physical Review Letters}, vol.~99, p.~156804, Oct. 2007.
\newblock Publisher: American Physical Society.

\bibitem{yang_fabry-perot_2020}
W.~Yang, C.~Urgell, S.~L. De~Bonis, M.~Marga{\'n}ska, M.~Grifoni, and A.~Bachtold, ``Fabry-{P}érot {Oscillations} in {Correlated} {Carbon} {Nanotubes},'' {\em Physical Review Letters}, vol.~125, p.~187701, Oct. 2020.
\newblock Publisher: American Physical Society.

\bibitem{pillet_andreev_2010}
J.-D. Pillet, C.~H.~L. Quay, P.~Morfin, C.~Bena, A.~Levy~Yeyati, and P.~Joyez, ``Andreev bound states in supercurrent-carrying carbon nanotubes revealed,'' {\em Nat. Phys.}, vol.~6, no.~12, pp.~965--969, 2010.
\newblock Publisher: Nature Publishing Group.

\bibitem{pillet_tunneling_2011}
J.-D. Pillet, {\em Tunneling spectroscopy of the {Andreev} {Bound} {States} in a {Carbone} {Nanotube}}.
\newblock phdthesis, Universit{\'e} Pierre et Marie Curie - Paris VI, Dec. 2011.

\bibitem{pillet_tunneling_2013}
J.-D. Pillet, P.~Joyez, R.~{\v Z}itko, and M.~F. Goffman, ``Tunneling spectroscopy of a single quantum dot coupled to a superconductor: {From} {Kondo} ridge to {Andreev} bound states,'' {\em Phys. Rev. B}, vol.~88, p.~045101, July 2013.

\bibitem{vion_thermal_1996}
D.~Vion, M.~G{\"o}tz, P.~Joyez, D.~Esteve, and M.~H. Devoret, ``Thermal {Activation} above a {Dissipation} {Barrier}: {Switching} of a {Small} {Josephson} {Junction},'' {\em Physical Review Letters}, vol.~77, pp.~3435--3438, Oct. 1996.
\newblock Publisher: American Physical Society.

\bibitem{jorgensen_critical_2007}
H.~I. J{\o}rgensen, T.~Novotn{\'y}, K.~Grove-Rasmussen, K.~Flensberg, and P.~E. Lindelof, ``Critical {Current} 0-$\pi$ {Transition} in {Designed} {Josephson} {Quantum} {Dot} {Junctions},'' {\em Nano Letters}, vol.~7, pp.~2441--2445, Aug. 2007.
\newblock Publisher: American Chemical Society.

\bibitem{eichler_tuning_2009}
A.~Eichler, R.~Deblock, M.~Weiss, C.~Karrasch, V.~Meden, C.~Sch{\"o}nenberger, and H.~Bouchiat, ``Tuning the {Josephson} current in carbon nanotubes with the {Kondo} effect,'' {\em Physical Review B}, vol.~79, p.~161407, Apr. 2009.
\newblock Publisher: American Physical Society.

\bibitem{Bretheau_supercurrent_spectro_2013}
L.~Bretheau, {\c{C}}.~{\"{O}}. Girit, C.~Urbina, D.~Esteve, and H.~Pothier, ``{Supercurrent Spectroscopy of Andreev States},'' {\em Physical Review X}, vol.~3, p.~041034, dec 2013.

\bibitem{feuillet-palma_conserved_2010}
C.~Feuillet-Palma, T.~Delattre, P.~Morfin, J.-M. Berroir, G.~F{\`e}ve, D.~C. Glattli, B.~Pla{\c c}ais, A.~Cottet, and T.~Kontos, ``Conserved spin and orbital phase along carbon nanotubes connected with multiple ferromagnetic contacts,'' {\em Physical Review B}, vol.~81, p.~115414, Mar. 2010.
\newblock Publisher: American Physical Society.

\bibitem{herrmann_carbon_2010}
L.~G. Herrmann, F.~Portier, P.~Roche, A.~L. Yeyati, T.~Kontos, and C.~Strunk, ``Carbon {Nanotubes} as {Cooper}-{Pair} {Beam} {Splitters},'' {\em Physical Review Letters}, vol.~104, p.~026801, Jan. 2010.

\bibitem{desjardins_synthetic_2019}
M.~M. Desjardins, L.~C. Contamin, M.~R. Delbecq, M.~C. Dartiailh, L.~E. Bruhat, T.~Cubaynes, J.~J. Viennot, F.~Mallet, S.~Rohart, A.~Thiaville, A.~Cottet, and T.~Kontos, ``Synthetic spin{\textendash}orbit interaction for {Majorana} devices,'' {\em Nature Materials}, vol.~18, pp.~1060--1064, Oct. 2019.
\newblock Publisher: Nature Publishing Group.

\bibitem{bordoloi_spin_2022}
A.~Bordoloi, V.~Zannier, L.~Sorba, C.~Sch{\"o}nenberger, and A.~Baumgartner, ``Spin cross-correlation experiments in an electron entangler,'' {\em Nature}, vol.~612, pp.~454--458, Dec. 2022.
\newblock Publisher: Nature Publishing Group.

\bibitem{wang_singlet_2022}
G.~Wang, T.~Dvir, G.~P. Mazur, C.-X. Liu, N.~van Loo, S.~L.~D. ten Haaf, A.~Bordin, S.~Gazibegovic, G.~Badawy, E.~P. A.~M. Bakkers, M.~Wimmer, and L.~P. Kouwenhoven, ``Singlet and triplet {Cooper} pair splitting in hybrid superconducting nanowires,'' {\em Nature}, vol.~612, pp.~448--453, Dec. 2022.
\newblock Publisher: Nature Publishing Group.

\bibitem{masseroni_spin-orbit_2024}
M.~Masseroni, M.~Gull, A.~Panigrahi, N.~Jacobsen, F.~Fischer, C.~Tong, J.~D. Gerber, M.~Niese, T.~Taniguchi, K.~Watanabe, L.~Levitov, T.~Ihn, K.~Ensslin, and H.~Duprez, ``Spin-orbit proximity in {MoS}$_2$/bilayer graphene heterostructures,'' Mar. 2024.
\newblock arXiv:2403.17120 [cond-mat].

\bibitem{dvir_realization_2023}
T.~Dvir, G.~Wang, N.~van Loo, C.-X. Liu, G.~P. Mazur, A.~Bordin, S.~L.~D. ten Haaf, J.-Y. Wang, D.~van Driel, F.~Zatelli, X.~Li, F.~K. Malinowski, S.~Gazibegovic, G.~Badawy, E.~P. A.~M. Bakkers, M.~Wimmer, and L.~P. Kouwenhoven, ``Realization of a minimal {Kitaev} chain in coupled quantum dots,'' {\em Nature}, vol.~614, pp.~445--450, Feb. 2023.
\newblock Publisher: Nature Publishing Group.

\bibitem{johannsen_fermionic_2024}
P.~D. Johannsen and C.~Schrade, ``Fermionic {Quantum} {Simulation} on {Andreev} {Bound} {State} {Superlattices},'' Apr. 2024.
\newblock arXiv:2404.12430 [cond-mat].

\bibitem{zhou_pressure_2024}
X.~Zhou, Y.~Chen, J.~Chen, C.~Hu, B.~Lyu, K.~Xu, S.~Lou, P.~Shen, S.~Ma, Z.~Wu, Y.~Xie, Z.~Zhang, Z.~L\"u, W.~Luo, Q.~Liang, L.~Xian, G.~Zhang, and Z.~Shi, ``Pressure-induced flat bands in one-dimensional moir\'e superlattices of collapsed chiral carbon nanotubes,'' {\em Phys. Rev. B}, vol.~109, p.~045105, Jan 2024.

\bibitem{Meng_self_consistent_2009}
T.~Meng, S.~Florens, and P.~Simon, ``{Self-consistent description of Andreev bound states in Josephson quantum dot devices},'' {\em Physical Review B}, vol.~79, p.~224521, jun 2009.

\bibitem{Maurand_First_2012}
R.~Maurand, T.~Meng, E.~Bonet, S.~Florens, L.~Marty, and W.~Wernsdorfer, ``First-order $0\mathrm{\text{\ensuremath{-}}}\ensuremath{\pi}$ quantum phase transition in the kondo regime of a superconducting carbon-nanotube quantum dot,'' {\em Phys. Rev. X}, vol.~2, p.~011009, Feb 2012.

\bibitem{kataura_optical_1999}
H.~Kataura, Y.~Kumazawa, Y.~Maniwa, I.~Umezu, S.~Suzuki, Y.~Ohtsuka, and Y.~Achiba, ``Optical properties of single-wall carbon nanotubes,'' {\em Synthetic Metals}, vol.~103, pp.~2555--2558, June 1999.

\bibitem{liu_atlas_2012}
K.~Liu, J.~Deslippe, F.~Xiao, R.~B. Capaz, X.~Hong, S.~Aloni, A.~Zettl, W.~Wang, X.~Bai, S.~G. Louie, E.~Wang, and F.~Wang, ``An atlas of carbon nanotube optical transitions,'' {\em Nature Nanotechnology}, vol.~7, pp.~325--329, May 2012.
\newblock Publisher: Nature Publishing Group.

\bibitem{Zorin1998}
A.~B. Zorin, ``{The thermocoax cable as the microwave frequency filter for single electron circuits},'' {\em Review of Scientific Instruments}, vol.~66, pp.~4296--4300, jun 1995.

\bibitem{lemasne_asymmetric_2010}
Q.~Le~Masne, {\em {Asymmetric current fluctuations and Andreev states probed with a Josephson junction}}.
\newblock Theses, {Universit{\'e} Pierre et Marie Curie - Paris VI}, Oct. 2010.

\bibitem{Goffman_Supercurrent_2000}
M.~F. Goffman, R.~Cron, A.~Levy~Yeyati, P.~Joyez, M.~H. Devoret, D.~Esteve, and C.~Urbina, ``Supercurrent in atomic point contacts and andreev states,'' {\em Phys. Rev. Lett.}, vol.~85, pp.~170--173, Jul 2000.

\bibitem{cron_these_2001}
R.~Cron, {\em {Les contacts atomiques : un banc d'essai pour la physique m{\'e}soscopique}}.
\newblock Theses, {Universit{\'e} Pierre et Marie Curie - Paris VI}, Nov. 2001.

\bibitem{tinkham_introduction_1996}
M.~Tinkham, {\em {INTRODUCTION} {TO} {SUPERCONDUCTIVITY} : {Second} edition}.
\newblock Dover {Books} on {Physics}, Mineola NY: Dover publications, 2nd ed.~ed., 1996.

\end{thebibliography}

\end{document}